# Template-assisted scalable nanowire networks

Martin Friedl[1], Kris Cerveny[2], Pirmin Weigele[2], Gozde Tütüncüoglu[1], Sara Martí-Sánchez[3], Chunyi Huang[4], Taras Patlatiuk[2], Heidi Potts[1], Zhiyuan Sun[4], Megan O. Hill[4], Lucas Güniat[1], Wonjong Kim[1], Mahdi Zamani[1], Vladimir G. Dubrovskii[5], Jordi Arbiol[3,6], Lincoln J. Lauhon[4], Dominik Zumbuhl[2], Anna Fontcuberta i Morral[1]

[1]*Laboratoire des Matériaux Semiconducteurs, École Polytechnique Fédérale de Lausanne, EPFL, 1015 Lausanne, Switzerland*

[2]*Department of Physics, University of Basel, Klingelbergstrasse 82, CH-4056 Basel, Switzerland*

[3]*Catalan Institute of Nanoscience and Nanotechnology (ICN2), CSIC and BIST, Campus UAB, Bellaterra, 08193 Barcelona, Catalonia, Spain*

[4]*Department of Materials Science and Engineering, Northwestern University, Evanston, Illinois 60208, United States*

[5]*ITMO University, Kronverkskiy pr. 49, 197101 St. Petersburg, Russia*

[6]*ICREA, Pg. Lluís Companys 23, 08010 Barcelona, Catalonia, Spain*

**Abstract**

Topological qubits based on Majorana fermions have the potential to revolutionize the emerging field of quantum computing by making information processing significantly more robust to decoherence. Nanowires (NWs) are a promising medium for hosting these kinds of qubits, though branched NWs are needed to perform qubit manipulations. Here we report gold-free templated growth of III-V NWs by molecular beam epitaxy using an approach that enables patternable and highly regular branched NW arrays on a far greater scale than what has been reported thus far. Our approach relies on the lattice-mismatched growth of InAs on top of defect-free GaAs nanomembranes (NMs) yielding laterally-oriented, low-defect InAs and InGaAs NWs whose shapes are determined by surface and strain energy minimization. By controlling NM width and growth time, we demonstrate the formation of compositionally graded NWs with cross-sections less than 50 nm.

Scaling the NWs below 20 nm leads to the formation of homogenous InGaAs NWs which exhibit phase-coherent, quasi-1D quantum transport as shown by magnetoconductance measurements. These results are an important advance towards scalable topological quantum computing.

**Keywords**: InAs, nanowires, GaAs, nanoscale membranes, template-assisted, weak-localization

**Introduction**

In the past few years, much progress has been made toward fabricating and scaling up qubit density to build universal quantum computing systems that can outperform classical computers by quantum schemes.[1–5] The ideal qubit should combine long coherence times, fast qubit manipulation, and small size, while maintaining scalability to many-qubit systems. Long coherence times are fundamentally challenging to achieve in various qubit systems due to the presence of numerous forms of environmental noise, requiring operating temperatures in the range of a hundred millikelvin.[6,7] A system which has been proposed to be much more robust against such perturbations is the topological qubit.[8,9] This type of qubit, for example, composed of Majorana fermions (MFs)[8,10] or parafermions (MPFs),[11,12] would have the inherent property of being topologically protected and would thus exhibit exceptionally long coherence times. Signatures of MF states have been observed experimentally in, among other systems, III-V semiconductor NWs in close proximity to an s-wave superconductor[13–15] while a few other groups have reported anomalous MF signatures in similar systems.[16–18] In general, these studies have focused on using III-V materials, such as InAs and InSb, due to their high spin-orbit coupling strength and g-factor.[19] Current efforts are focused on performing the first manipulations of MFs to further verify theoretical predictions, for which low-disorder, connected one-dimensional (1D) branches are required.[20] Gold-free and defect-free NW branches made of a high-purity, high-spin-orbit III-V material would be an ideal platform for manipulating MFs. Excellent progress has been made towards this goal with reports on the growth of monocrystalline gold-assisted InSb NW branches which display weak anti-localization due to the large spin-orbit interaction of the material, as well as a hard superconducting gap.[21,22] Scalability is another important aspect of any future computational system and, on this front, the Riel group has recently demonstrated patternable ballistic InAs NW crosses through template-assisted growth on silicon.[23]

Despite recent progress, a few challenges still exist with current methods to produce branched structures. The fabrication of NW networks and intersections has been explored for many years for classical computing by overlapping individual wires.[24–26] For MF applications, the stringent requirement of maintaining coherent transport across the intersection means that, currently, the most popular NW cross structures rely on the intersection of two gold-catalysed NWs grown along two different <111>B directions leading to an interface-free junction.[27–31] After growth, free-standing crosses are obtained which then need to be transferred onto a separate substrate for further device fabrication, limiting the ultimate scalability. A scalable scheme would instead enable the NW growth and intersections to be realized directly on the final device substrate. At the same time, for future device integration, the use of gold seeds poses a problem for compatibility with CMOS technologies.[32] Here, we demonstrate a new approach to grow gold-free branched In(Ga)As NWs at the wafer scale by using GaAs NMs as templates.

Defect-free GaAs NMs of exceptional quality constitute the ideal templates for further In(Ga)As NW growth.[33,34] Such structures have been successfully grown by both metalorganic chemical vapour deposition (MOCVD) and molecular beam epitaxy (MBE) using a gold-free selective area approach.[35,36] The NMs are patternable at the wafer scale and can additionally be fabricated in the form of Y-shaped structures by growing them along the three <-1-12> directions on GaAs (111)B substrates.[35] When the growth of these GaAs NMs is followed by InAs, the InAs accumulates at the top of the NMs, forming In(Ga)As NWs along the NM vertex, as depicted schematically in Figure 1a. Shown in Figure 1b is the progression of the NW growth which initiates as InGaAs and then evolves to pure InAs for longer growth times. By varying the deposition times and growth conditions, the size and composition of the NWs can be changed. Combining the concepts of patterning NMs into Y-branches and performing In(Ga)As NW growth on top of GaAs NMs, Y-shaped In(Ga)As NW junctions can be obtained, as shown in Figure 1c and Figure 1d. Our approach thereby enables the growth of gold-free branched NWs at the wafer scale.

In this paper, as the first step toward building MF devices based on this approach, we demonstrate the growth of low-defect linear In(Ga)As NWs based on GaAs NMs. Magnetotransport measurements (depicted schematically in Figure 1e) demonstrate weak localization in the diffusive regime suggesting quasi-1D quantum transport thus making such NWs candidates for future quantum computing schemes.

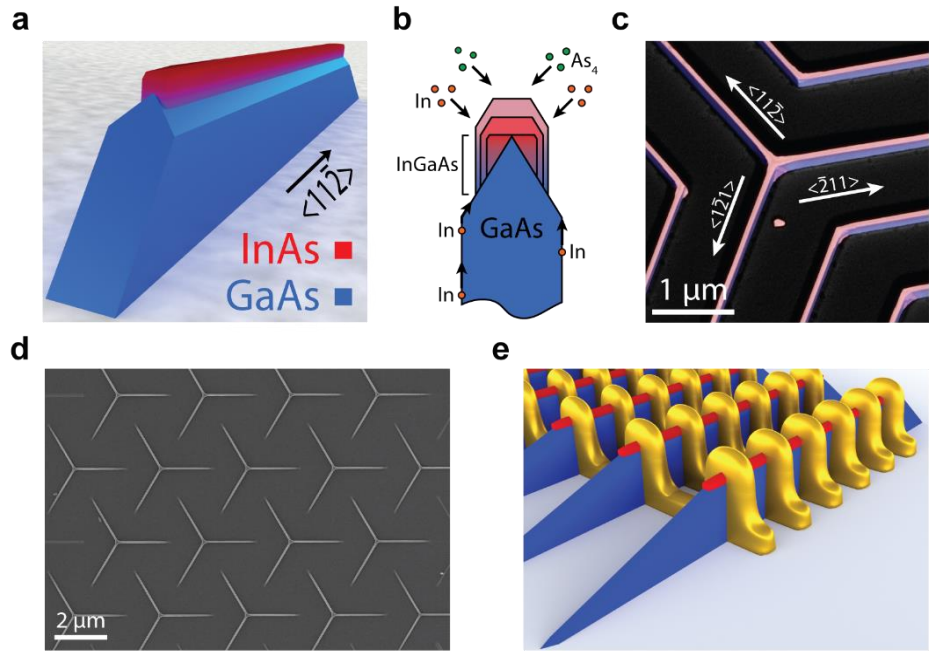

**Figure 1 | Growth of NWs on NMs.** (**a**) Model of a single GaAs/InAs NM/NW structure. (**b**) Diagram showing NW growth progression. (**c**) 30° tilted scanning electron microscope (SEM) image of a branched NM/NW structure using an energy-selective backscattered electron detector for z-contrast. False-coloured for visibility and annotated with relevant substrate directions. (**d**) SEM image of branched NW structures grown in a regular array. (**e**) Model of an array of contacted linear NM/NW structures used in magnetotransport measurements.

**Nanomembranes as a Platform**

Membranes with top ridges parallel to the substrate were grown as described previously.[36] The InAs NWs were then grown for 200s at an $As_4$ flux of $8\times10^{-6}$ Torr, an In rate of 0.2 Å/s and a substrate temperature of 540°C. This optimized substrate temperature yielded long, continuous InAs growth on top of the NMs. The details can be found in the supplementary information (SI). A Si dopant flux of $\sim10^{13}$ atoms/cm$^2$/s was also introduced to increase the conductivity of the NWs.

The NW morphology, composition, and structural quality were extensively characterized by correlated analysis using various electron microscopy techniques. These included electron energy loss spectroscopy (EELS) and atomic resolution aberration-corrected annular dark field scanning transmission electron microscopy (ADF-STEM). The results were then coupled with geometrical phase analysis (GPA) to give strain information,[37,38] which in turn was fed into a semi-empirical model to understand the formation of the NWs from a theoretical standpoint.

Analysis of focused ion-beam (FIB) lamellas prepared perpendicular to the NW axis by ADF-STEM and correlated EELS (Figure 2a-c) show that ~50 nm diameter InAs NWs form on the 250 nm tall GaAs NMs. The InAs material preferentially accumulates along the top ridge of the GaAs NM, forming the NW which is primarily InAs with a ~20 nm thick intermixed InGaAs region at its base (see SI for details). This InGaAs region likely occurs due to strain-mediated intermixing with the GaAs NM below as has been observed in InAs quantum dots on GaAs.[39] It is important to note that we observe no In signal from the NMs; the faint signal seen in the EELS map is believed to be created during the FIB cutting by a combination of redeposition of the $TiO_x$ protective layer and surface diffusion of the highly mobile In adatoms. Looking instead at the NW facets, as seen in Figure 2d, the resulting InAs NW structures are terminated by two (110) facets on the sides and have a single flat (111) top facet. The appearance of this (111) facet, instead of the two {113} facets as in the GaAs NMs, can be explained by the higher $As_4$ flux used in the InAs NW growth.[40,41]

No defects were observed when viewing this transverse lamella in atomic-resolution ADF-STEM mode. As strain along the NW axis was predicted to be more difficult to relax than in the transverse direction, a second FIB lamella was prepared parallel to the axis of the NW and also imaged using atomic-resolution ADF-STEM. Here, a few misfit dislocations were observed to near the InAs/GaAs interface, with an estimated density of approximately 100/µm, as described in the SI. This constitutes a 40% reduction in dislocation density with respect to the equivalent 2D growth of InAs on GaAs and at least three times improvement with respect to the twin density typically observed in self-catalyzed InAs NWs.[42–44]

Turning now to explaining the morphology of the structures, a simple analytical model shows that the surface and strain energy minimization play the most important role in driving the NW to adopt the observed shape (see SI). Taking advantage of the atomic resolution offered by ADF-STEM images, GPA was performed on the initial FIB lamella cut perpendicular to the NW axis. Looking specifically at the (111) plane spacing, a substantial 2-3% residual compressive strain is observed within the NW, as shown in Figure 2e, with the corresponding line scan given in Figure 2f. Using this strain value, a semi-quantitative model was developed to describe the total NW energy of formation, taking into account the InAs/GaAs surface energies and the InAs elastic strain energy. The total energy of the system was then minimized with respect to the NW aspect ratio. Interestingly, the experimentally observed aspect ratio coincides with that of the theoretical minimum energy shape, suggesting that the NW shape is simply driven by energy minimization (see SI for details).

**Electrical Transport in the Mesoscopic Regime**

To bring the NWs into the 1D electrical transport regime, they were downscaled to ~20 nm diameters by narrowing the GaAs NMs by using smaller $SiO_2$ openings and shorter growth times for less lateral growth. The resulting NWs were smaller both laterally and also vertically, resulting in intermixed InGaAs NWs, as depicted for small diameter wires in the growth progression diagram in Figure 1b. These results were confirmed by performing atom probe tomography (APT) which additionally yielded information about the Si dopant distribution.

The APT analysis confirmed the presence of an InGaAs NW while also uncovering signs of dopant rejection during crystal growth, causing an accumulation of Si atoms at the NW surface. Figure 2g shows a typical APT map of the In mole fraction and Si dopant distribution. A quantitative composition profile of the NW surface was extracted in the proximity histogram (proxigram) shown in Figure 2h. In this sample, the NW group III mole fractions are ~17% In and ~83% Ga (additional maps are given in the SI). Analysis of different InGaAs samples by APT tomography under similar conditions shows a slight tendency towards preferential In evaporation. In addition, a GaAs capping layer was deposited on the APT sample after the NW growth which may have enhanced the Ga intermixing.[45] For these reasons, we consider the In mole fraction in this NW as a lower bound. Although a Si flux was present during InAs NW growth, the Si atoms are not homogenously distributed throughout the NW (Figure 2g). The Si atoms instead appear to accumulate preferentially at the (111) growth interface, resulting in a quasi-remotely-doped structure.

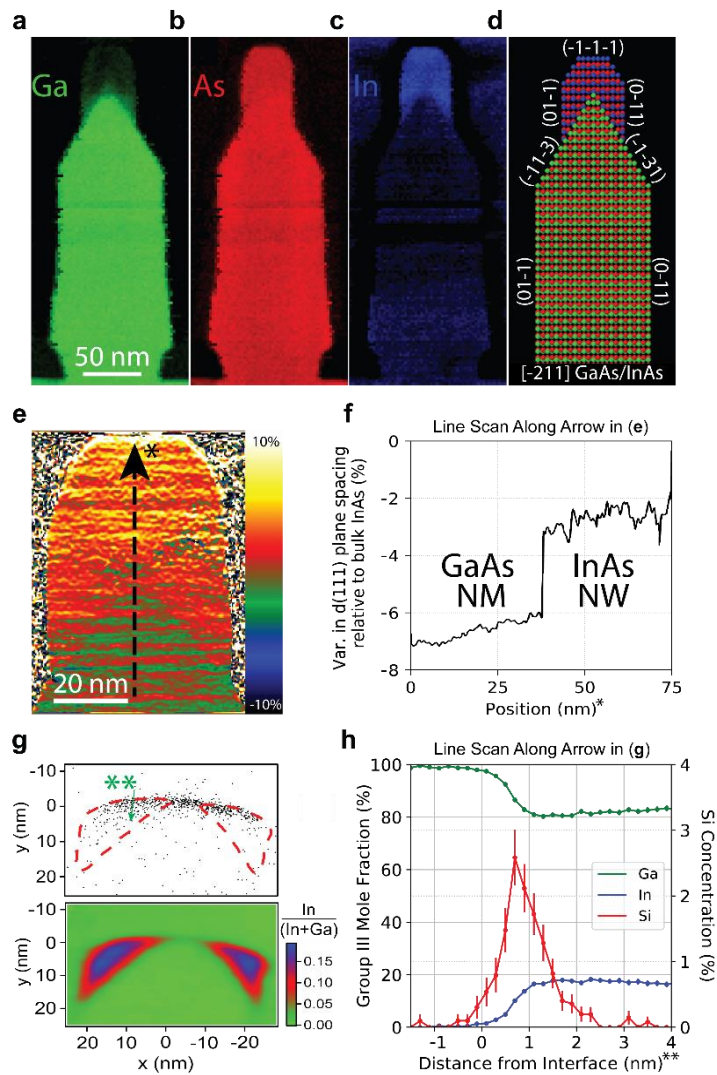

**Figure 2 | TEM and APT Analysis.** (**a-c**) EELS maps with elemental contrast of NM/NW cross-section. Note that the overlap of the In EELS signal with the Ti signal from the $TiO_x$ capping layer has caused an anomalous background of In within the NM which is not physical. (**d**) Atomic model showing faceting of the NM/NW heterostructure. (**e**) GPA map of the InAs NW region. (**f**) Line scan along arrow in **e**. (**g**) APT map of scaled-down NWs used in electrical measurements showing In concentration (lower map) and accumulation of Si dopant atoms (black dots) at surface of the NW with In isoconcentration lines as a guide to the eye (upper map). Note: since the NWs were capped with GaAs for APT analysis, the measured In concentration can be considered as a lower bound for the uncapped structures. (**h**) Proxigram line scan along dashed arrow in **g** showing Si accumulation on the NW top facet and In concentration within the NW.

The electrical properties of the NWs were explored through multi-contact resistance measurements on an array of NWs for which an example device is shown in Figure 3a. An array of 34 NWs, comparable to those on which APT was performed, was used for these tests as a way to obtain the average response from many devices. Standard four-point measurements were then carried out at room and low temperatures before moving to low-temperature magnetoconductance transport experiments.

Room temperature transmission line measurements, Figure 3b, gave linearly scaling and repeatable resistances, suggesting that a good-quality contact was achieved. A control sample without InAs NWs (as shown in green in Figure 3b) shows a five orders of magnitude increase in resistance, directly proving that the observed conduction occurs due to the InAs deposition. Device behaviour remained linear and ohmic down to 4.2K, albeit with an increased contact resistance.

Magnetoconductance measurements at 1.5 K revealed a zero-field minimum of conductance consistent with weak localization (WL) behaviour, as shown in Figure **3**c.[46] The conductance of the devices was measured under constant bias while the magnetic field was applied perpendicular to the substrate and was swept from -8T to 8T. Analysis indicates conduction in the quasi-1D transport regime and elucidates important quantum figures of merit such as coherence length and spin-orbit length, as described below.

A coherence length, $l_\varphi$, of 130 nm and a lower bound on the spin-orbit length, $l_{so}$, of 280 nm were obtained by fitting of the experimental data. The system was assumed to be in the diffusive regime with the electron mean free path, $l_e \ll W$, with $W$ being the width of the conducting channel, estimated from the APT results to be about 20 nm. In this regime $l_e$ is thus constrained to be a few nanometres due to the large amount of dopant and surface scattering. A simple quasi-1D model for the quantum corrections to the conductivity in the diffusive limit (see methods) gives excellent agreement with the data and yields $l_\varphi \sim 130 \pm 4$ nm near zero bias, as shown in the solid traces in Figure 3c. In the absence of weak anti-localization, and adding spin-orbit coupling to the model, a lower bound for $l_{so}$ of 280 nm is estimated. Some small variations in the conductance at large magnetic fields are noticeable, presumably signatures of conductance fluctuations, which are strongly suppressed here due to averaging from the parallel NW arrangement as well as the relatively short coherence length.

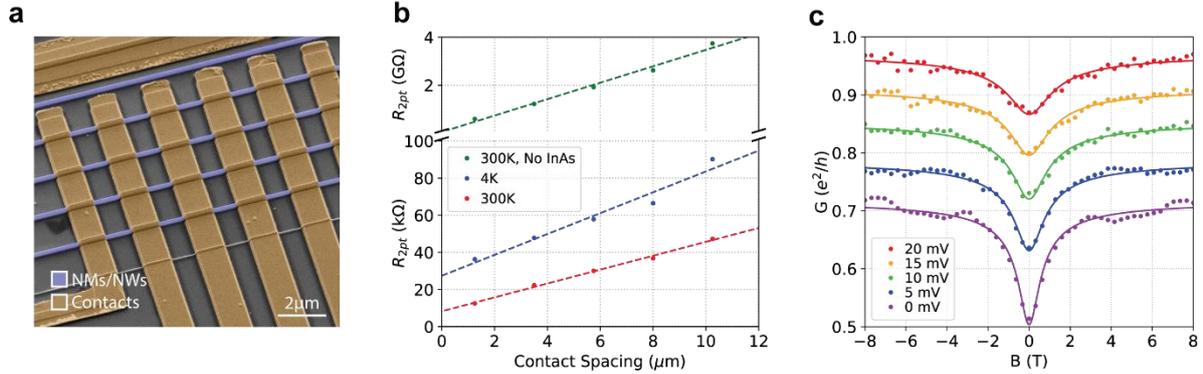

**Figure 3 | Magnetotransport Measurements.** (**a**) False-coloured SEM image of four InAs/GaAs NW/NMs contacted in parallel. (**b**) Transmission line measurement to extract contact resistance and resistivity (per NW) at 4K and 300K, in comparison with NMs without the InAs deposition step. (**c**) Average differential conductance per NW as a function of magnetic field perpendicular to the NWs for a range of bias voltages measured at 1.5K for a 1.25 µm-long NW segment. The traces for biases above 0 mV are offset for clarity.

These initial results show that electron confinement to the top of NM-grown NWs is sufficient to produce quasi-1D conduction. These NWs could therefore be viable hosts for Majorana bound states provided two potential obstacles are addressed. First, a NW made of higher spin-orbit material such as pure or nearly-pure InAs or InSb will be required, i.e. intermixing should be reduced or eliminated. This will be apparent by the observation of weak anti-localization instead of weak-localization. Second, impurity, interface and alloy scattering need to be reduced such that $l_e$ transitions from the diffusive to the ballistic transport regime.

**Conclusion & Outlook**

The fast-growing field of quantum computing and the promise of robust, topologically-protected qubits with III-V NWs drives the pursuit of scalable approaches to branched NW networks. We have described a path forward using GaAs NMs as templates for In(Ga)As NW growth. By exploiting strain in the highly mismatched InAs/GaAs system, continuous, low-defect NWs were formed. We have further observed weak localization, demonstrating that such NWs can provide sufficient confinement to achieve quasi-1D conduction. Our gold-free wafer-scale approach to branched NWs serves as a platform for future investigations into 1D transport and quantum computation with III-V NW networks with many exciting possibilities. From an MBE growth perspective, using GaSb NMs, already described by other groups, to grow InSb NWs would be interesting due to the higher g-factor of InSb.[19,47] Alternatively, suitable plastic strain relaxation, for example by interfacial misfit array formation,[48] may enable GaAs NMs to be viable templates for InSb NW growth. At the same time, the growth of new kinds of structures with additional functionalities is another avenue to explore, including for example research into parafermion devices by stacking multiple NWs on top of eachother.[12] The wealth of intriguing new possibilities and approaches offered by template-assisted III-V NW growth makes this method an important step towards one day realizing a scalable quantum computing scheme based on NW topological qubits.

**Methods**

**Substrate Preparation.** Undoped GaAs (111)B substrates were prepared by first depositing 25 nm of $SiO_2$ by plasma enhanced chemical vapour deposition (PECVD). This was followed by e-beam lithography using ZEP resist and low-temperature development to achieve low line edge roughness.[49] Subsequent dry etching with fluorine chemistry was used to etch the $SiO_2$ down to the GaAs surface and a final wet etch in a dilute buffered HF solution helped remove any remaining oxide. This yielded openings varying from 30 nm to 100 nm in width and 10-20 μm in length, depending on the e-beam pattern.

**Growth.** The nanostructures were grown in a DCA D600 Gen II solid-source MBE. The optimal growth of the GaAs/InAs NM/NW heterostructures was found to be at a temperature of 630°C/540°C (as measured by the pyrometer), As flux of $4\times10^{-6}/8\times10^{-6}$ Torr and Ga/In growth rates of 1.0/0.2 Å/s, respectively. The NMs were typically grown for 30 min (180 nm nominal 2D thickness) while the InAs NWs were grown for 200s (4 nm nominal 2D thickness).

**(S)TEM.** The cross sections of the NMs were prepared by focused ion beam (FIB) milling normal to the substrate surface and investigated by atomic resolution aberration corrected annular dark field scanning transmission electron microscopy (ADF-STEM) in a probe corrected FEI Titan 60-300 keV microscope operated at 300 keV. The elemental maps were obtained by using an Electron Energy Loss Spectrometer (EELS) coupled to a Tecnai F20 microscope.

**Contacting.** Contacts were patterned by e-beam lithography followed by dual-angle evaporation of 14/80 nm of Cr/Au for good side-wall coverage. Before metallization, an O$_2$ plasma clean and a 6 min ammonium polysulfide etch at 40°C were used to ensure a clean, oxide-free contact.[50]

**Quantum Transport Model.** The conductance of the NW is described as:

$$\Delta G = -\frac{2e^2}{hL}\left(\frac{1}{l_\varphi^2}+\frac{1}{l_B^2}\right)^{-\frac{1}{2}},$$

where $L$ is the spacing between the contacts and $l_B$ is the magnetic dephasing length $l_B = \sqrt{D\tau_B}$ with $D$ as the diffusion constant ($D_{1D} = v_F l_e$ for 1D diffusion). In this limit, the magnetic dephasing time $\tau_B$ is given as $\tau_B = 3l_m^4/W^2D$, where $l_m = \sqrt{\hbar/eB}$ is the magnetic length.[46,51,52]

**GaAs Capping.** After the growth of the NW/NM heterostructures, a ∼30 nm GaAs cap was deposited in-situ by MBE at 400°C. In the middle of this GaAs cap, the In shutter was opened for 10s, making a few-monolayer insertion of $In_{0.16}Ga_{0.84}As$ indicating the midpoint of the GaAs cap. The NW/NM heterostructures were then coated with a ∼110 nm GaAs layer[53] with ion beam sputtering, at 9 kV and 7.5 mA for 1 h. The capping layer protected the sample from damage caused by the ion beam during FIB.

**APT.** A standard lift-out method[53,54] was performed in a FEI Helios dual-beam focused ion-beam (FIB) microscope with a micromanipulator, and the as-prepared wedge-shaped samples were welded onto Si microposts. Finally, the needle-shape APT specimens were obtained by ion-beam annular milling. APT was performed with a local-electrode atom-probe (LEAP) 4000X Si tomograph (Cameca, Madison, WI) at a sample temperature of 40 K and a background pressure of $3\times10^{-11}$ Torr. An ultraviolet focused laser with wavelength of 355 nm was used to evaporate the sample atoms into ions, at a pulse rate of 250 kHz and detection rate of 0.7%. The pulse energy was gradually changed from 1.2pJ to 0.8pJ during the evaporation process. The data was reconstructed utilizing IVAS 3.8.1 to provide a 3D composition profile. SEM images of the nanotips taken in the FIB were used to guide the choices of the reconstruction parameters.

**Associated Content**

**Supporting Information.** Additional information on growth studies, strain analysis, growth model, NW composition, defect density, magnetotransport and APT mass spectra.

**Acknowledgements**

Authors from EPFL and U. Basel acknowledge funding through the NCCR QSIT. Authors from EPFL further thank funding from SNF (project no. IZLRZ2_163861) and H2020 via the ITN project INDEED. SMS acknowledges funding from "Programa Internacional de Becas "la Caixa"-Severo Ochoa". JA and SMS acknowledge funding from Generalitat de Catalunya 2017 SGR 327 and the Spanish MINECO coordinated project ValPEC (ENE2017-85087-C3). ICN2 acknowledges support from the Severo Ochoa Programme (MINECO, Grant no. SEV-2013-0295) and is funded by the CERCA Programme / Generalitat de Catalunya. This work has received funding from the European Union's Horizon 2020 Research and Innovation Programme under grant agreement No. 654360 NFFA-Europe. Part of the present work has been performed in the framework of Universitat Autònoma de Barcelona Materials Science PhD program. The atomic resolution ADF-STEM microscopy was conducted in the Laboratorio de Microscopias Avanzadas at the Instituto de Nanociencia de Aragon-Universidad de Zaragoza. JA and SMS thank them for offering access to their instruments and expertise. VGD thanks the Ministry of Education and Science of the Russian Federation for financial support under grant 14.613.21.0055 (project ID: RFMEFI61316X0055). LJL acknowledges support of NSF DMR-1611341. M.O.H. acknowledges support of the NSF GRFP. Atom-probe tomography was performed at the Northwestern University Center for Atom-Probe Tomography (NUCAPT). The LEAP tomograph at NUCAPT was purchased and upgraded with grants from the NSF-MRI (DMR-0420532) and ONR-DURIP (N00014-0400798, N00014-0610539, N00014-0910781, N00014-1712870) programs. NUCAPT received support from the MRSEC program (NSF DMR-1720139)

at the Materials Research Center, the SHyNE Resource (NSF ECCS-1542205), and the Initiative for Sustainability and Energy (ISEN) at Northwestern University.

**TOC Graphic**

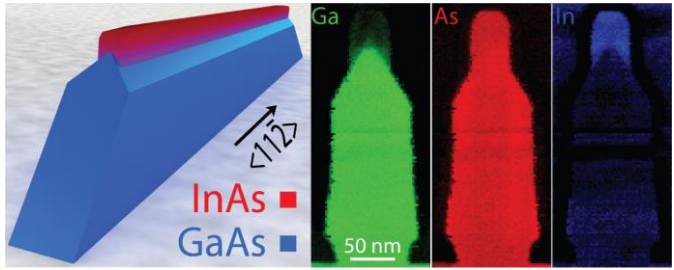

Template-assisted nanowire structure and cross-sectional STEM-EELS elemental mapping.